\documentclass[preprint,number,sort&compress]{elsarticle}
\usepackage{a4p}
\usepackage{hyperref}
\usepackage{subfigure}
\usepackage{siunitx}
\usepackage{atlasphysics}
\usepackage{amssymb}
\usepackage{amsmath}
\usepackage{color}
\newcommand{\dedx}{\ensuremath{\mathrm{d}E\!/\!\mathrm{d}x}}
\newcommand{\dedxp}{\ensuremath{\mathrm{d}E\!/\!\mathrm{d}x_{\rm Pixel}}}
\newcommand{\betat}{\ensuremath{ {\beta}_{\mathrm{Tile}}}}
\makeatletter
\DeclareRobustCommand*{\bfseries}{%
\not@math@alphabet\bfseries\mathbf
\fontseries\bfdefault\selectfont
\boldmath
}
\makeatother

\journal{Phys. Lett. B}

\begin{document}

\begin{frontmatter}



\title{Search for stable hadronising squarks and gluinos with the \mbox{ATLAS }experiment at the LHC\\
{\small CERN-PH-EP-2011-026}
}
\author{The ATLAS Collaboration}

\begin{abstract}
Hitherto unobserved long-lived massive particles with electric and/or colour charge are predicted by a range of theories which extend the Standard Model. In this paper a search is performed at the ATLAS experiment for slow-moving charged particles produced in proton-proton collisions at 7\,\si{\tera\electronvolt}{} centre-of-mass energy at the LHC, using a data-set corresponding to an integrated luminosity of 34\,pb$^{-1}$. No deviations from Standard Model expectations are found.
This result is interpreted in a framework of supersymmetry models in which coloured sparticles can hadronise into long-lived bound hadronic states, termed $R$-hadrons, and 95\% CL limits are set on the production cross-sections of squarks and gluinos. 
The influence of $R$-hadron interactions in matter was studied using a number of different models, and lower mass limits for stable sbottoms and stops are found to be 294 and 309\,\si{\giga\electronvolt}{} respectively. The lower mass limit for a stable gluino lies in the range from 562 to 586\,\si{\giga\electronvolt}{} depending on the model assumed. Each of these constraints is the most stringent to date. 
\end{abstract}


\end{frontmatter}


%
%
\section{Introduction}
\label{sec:introduction}
The discovery of exotic stable massive particles (SMPs)\footnote{The term stable is taken in this paper to mean that the particle has a decay length comparable to the size of the \mbox{ATLAS} detector or longer.} at the LHC would be of fundamental significance. The motivation for SMP searches at \mbox{ATLAS} arises, for example, from proposed solutions to the gauge hierarchy problem, which involve previously unseen particles with TeV-scale masses~\cite{Fairbairn:2006gg,Raklev:2009mg}. The ATLAS experiment has recently searched for SMPs with large electric charge~\cite{:2011mb}. SMPs possessing colour charge represent another class of exotic particle which can be sought. Hadronising SMPs are anticipated in a wide range of exotic physics models~\cite{Fairbairn:2006gg} that extend the Standard Model (SM). For example, these particles appear in both $R$-parity conserving supersymmetry (SUSY) and universal extra dimensions. The possibility of direct pair production through the strong nuclear force implies large production cross-sections. Searches for these particles are thus an important component of the early data exploitation programs of the LHC experiments~\cite{Aad:2009wy}. In this paper, the first limits from the \mbox{ATLAS }experiment are presented on the production of coloured, hadronising SMPs in proton-proton collisions at 7~TeV centre-of-mass energy at the LHC. Results are presented in the context of SUSY models predicting the existence of $R$-hadrons~\cite{Farrar:1978xj}, which are heavy objects formed from a coloured sparticle (squark or gluino) and light SM partons. 

SMPs produced at LHC energies typically possess the following characteristics: they are penetrating\footnote{A small fraction of SMPs can be brought to rest by interactions in the detector. Should they have finite lifetimes an alternative approach to the direct detection of SMPs would be to observe their decays~\cite{Khachatryan:2010uf}.} and propagate at a low enough speed that they can be observed as being subluminal using measurements of time-of-flight and specific ionisation energy loss \cite{Fairbairn:2006gg}.  Previous searches for $R$-hadrons have typically been based on either the signature of a highly ionising particle in an inner tracking system~\cite{Abreu:1998jy,Heister:2003hc,Collaboration:2011ts} or a slow-moving muon-like object~\cite{Abazov:2008qu,Aaltonen:2009ke,Collaboration:2011ts}. The latter limits rely on the assumption that the $R$-hadron is electrically charged when it leaves the calorimeter and can thus be detected in an outer muon system. However, hadronic scattering of $R$-hadrons in the dense calorimeter material, and the properties of different mass hierarchies for the $R$-hadrons, may render most of the produced $R$-hadrons electrically neutral in the muon system\cite{Mackeprang:2009ad}.  Such an effect is expected for $R$-hadrons formed from sbottom-like squarks~\cite{deBoer:2007ii}; the situation for gluino-based $R$-hadrons is unclear, with different \mbox{models} giving rise to different phenomenologies. The previous mass limit for gluino $R$-hadrons with minimal sensitivity to scattering uncertainties is 311\,\si{\giga\electronvolt}{} at 95\% confidence level~\cite{Collaboration:2011ts} from the CMS collaboration.

The \mbox{ATLAS }detector contains a number of subsystems which provide information which can be used to distinguish SMPs from particles moving at velocities close to the speed of light. Two complementary subsystems used in this work are the pixel detector, which measures ionisation energy loss (\dedx), and the tile calorimeter, which measures the time-of-flight from the interaction point for particles which traverse it. Furthermore, since there is no requirement that a candidate be reconstructed in the outer muon spectrometer, the search is robust to theoretical uncertainties on the fraction of $R$-hadrons that are charged when leaving the calorimeter system.  The analysis extends the mass limits beyond already published limits and represents the first dedicated direct search for sbottom $R$-hadrons at a hadron collider.
%
%
\section{Simulation of R-hadrons and background processes}\label{sec:mc}

Monte Carlo simulations are used primarily to determine the efficiency of the $R$-hadron selection together with the associated systematic uncertainties. Predicted backgrounds are estimated using data, as described in Section~\ref{sec:results}. However, simulated samples of background processes (QCD and $t \bar{t}$, $W$ and $Z$ production) are used to optimise the $R$-hadron selections, without biasing the selection in data. 

Pair production of $\tilde{g}\tilde{g}$, $\tilde{t}\bar{\tilde{t}}$ and $\tilde{b}\bar{\tilde{b}}$ is simulated in {\sc 
Pythia}~\cite{Sjostrand:2006za} using the DW tune~\cite{Albrow:2006rt,Sjostrand:1987su}. The string hadronisation model~\cite{Andersson:1983ia}, incorporating specialised hadronisation routines~\cite{Fairbairn:2006gg} is used to produce final states containing $R$-hadrons. For gluino scenarios the probability for a gluino to form a gluon-gluino bound state, based on a colour octet model, is assumed to be 10\%~\cite{Fairbairn:2006gg}. The simulation of $R$-hadron interactions in matter is handled by dedicated {\sc Geant4} routines \cite{Agostinelli:2002hh,:2010wqa} based on three different models with alternative assumptions. $R$-hadrons containing squarks are simulated using the model described in Ref.~\cite{deBoer:2007ii}. This model is motivated by extrapolations from SM heavy quark hadron spectra. It furthermore employs a triple-Regge formalism to describe hadronic scattering. For gluino $R$-hadrons there are less strict theoretical constraints since no SM analogue exists for a heavy colour octet. Consequently a physics model is chosen, as described in Refs.~\cite{Kraan:2004tz,Mackeprang:2006gx}. This model has been used in other publications~\cite{Kraan:2005ji,Khachatryan:2010uf,Collaboration:2011ts} and it imposes few constraints on allowed stable states. Doubly charged $R$-hadrons and a wide variety of charge reversal signatures in the detector are possible. Hadronic scattering is described through a purely phase space driven approach. More recent models for the hadronic scattering of gluino $R$-hadrons predict that the majority of all produced $R$-hadrons will be electrically neutral after just a few hadronic interactions. One of these models is an extension of the triple-Regge model used to describe squark $R$-hadrons~\cite{Mackeprang:2009ad}. Another is the bag-model based calculation presented in Ref.~\cite{Farrar:2010ps}. Independent results for gluino $R$-hadrons are presented here for these models.

The simulated samples have gluino (squark) masses in the range 100-700\,\si{\giga\electronvolt}{} (100-500\,\si{\giga\electronvolt}), roughly matching the sensitivity that can be achieved given the statistical precision of the data sample on which the present analysis is based.  The cross-sections of the individual samples are normalised to the predictions of the {\sc Prospino} NLO program~\cite{Beenakker:1996ch}
using CTEQ 6.6 parton density functions (PDFs) ~\cite{Nadolsky:2008zw}. All other sparticles are set to high mass and are 
decoupled from the calculations used in this work.

\section{The \mbox{ATLAS }detector}
The \mbox{ATLAS }detector is described in detail in Ref.~\cite{:2008zzm}. Below, some features of the subsystems most important for the present analysis are outlined.

\subsection{Specific energy loss from the  pixel detector}
\label{sec:pixdet}
As the innermost sub-detector in ATLAS, the silicon-based pixel detector contributes to precision tracking in the region\footnote{ATLAS uses a right-handed coordinate system with its origin at the nominal interaction point (IP) in the centre of the detector and the $z$-axis along the beam pipe. The $x$-axis points from the IP to the centre of the LHC ring, and the $y$ axis points upward. Cylindrical coordinates $(r,\phi)$ are used in the transverse plane, $\phi$ being the azimuthal angle around the beam pipe. The pseudorapidity is defined in terms of the polar angle $\theta$ as $\eta=-\ln\tan(\theta/2)$.} $|\eta| <2.5$. The sensitive detectors of the pixel detector barrel are placed on three concentric cylinders around the beam-line, whereas each end-cap consists of three disks arranged perpendicular to the beam axis. The pixel detector therefore typically provides at least three measurements for each track. In the barrel (end-cap) the intrinsic accuracy is 10\,\si{\micro\metre} in the $r$-$\phi$ plane and 115\,\si{\micro\metre} in the $z$ ($r$)-direction. The integrated time during which a signal exceeds threshold has a sub-linear dependence on the charge deposited in each pixel. This has been measured in dedicated calibration scans, enabling an energy loss measurement for charged particles using the pixel detector.

The charge released by a track crossing the pixel detector is rarely contained within just one pixel. Neighbouring pixels are joined together to form clusters, and the charge of a cluster is calculated by summing up the charges of all pixels after applying a calibration correction. The specific energy loss, \dedx, is estimated as an average of the individual cluster \dedx{} measurements (charge collected in the cluster, corrected for the track length in the sensor), for the clusters associated with the track. To reduce the effects of the Landau tail, the \dedx{} of the track is calculated as the truncated mean of the individual cluster measurements. In the study presented here at least two clusters are required for the pixel detector {\dedx} measurement (\dedxp). Further details and performance of the method are described in~\cite{dedxnote}.

\subsection{Time-of-Flight from the tile calorimeter}
The \mbox{ATLAS} tile calorimeter is a sampling calorimeter that constitutes the barrel part of the hadronic calorimetry in \mbox{ATLAS}. It is situated in the region 2.3~$< r <$~4.3\,m, covering $|\eta| \lesssim 1.7$, and uses iron as the passive material and plastic scintillators as active layers. Along the beam axis, the tile calorimeter is logically subdivided into four partitions, each segmented in equal intervals of azimuthal angle ($\phi$) into 64 modules. The modules are further divided into cells, which are grouped radially in three layers, covering 0.1 units in $\eta$ in the first two layers and 0.2 in the third. Two bundles of wavelength-shifting fibres, associated with each cell, guide the scintillation light from the exposed sides of the module to photomultiplier tubes. The signal from each photomultiplier tube is digitised using dual ADCs covering different dynamic ranges. Analysing seven consecutive samplings with an interval of 25\,\si{\nano\second} allows the amplitude, pedestal value and peak position in time to be extracted.  The tile calorimeter provides a timing resolution of 1-2\,\si{\nano\second} per cell for energy deposits typical of minimum-ionising particles (MIPs). The measured times have been corrected for drifts in the LHC clock using high-precision timing measurements from a beam pick-up system~\cite{Ohm:2009pf} and calibrated such that energy depositions associated with muons from $Z$-boson decays are aligned at $t=0$ in both data and simulations.

Although the readout electronics have been optimised to provide the best possible timing resolution for \mbox{$\beta = 1$} particles, the performance for slower particles ($0.3 < \beta <1$) is not seriously compromised. In addition, SMPs tend to traverse the entire tile calorimeter, leaving statistically independent signals in up to six cells. 

The time-of-flight and hence the speed, $\beta$, of an $R$-hadron candidate can be deduced from time measurements in the tile calorimeter cells along the candidate trajectory. All cells along the particle trajectory with an energy deposition larger than 500~\si{\mega\electronvolt} are used to make an independent estimate of $\beta$. The time resolution has been shown to improve with the energy measured in the cell~\cite{Leitner:2007}, so the cells are combined using an average weighted by cell energy to get a velocity measurement (\betat). Combining the measurements from all cells results in a time resolution of $\sim$1\,\si{\nano\second}.

\section{Event selection}\label{sec:results}
The data sample used in this work corresponds to an integrated luminosity of 34\,pb$^{-1}$.
Final states with $R$-hadrons can also contain jets and missing transverse energy ($\MET$) arising from QCD radiation which can be used to select candidate events.  Due to the large cross-section for jet production at the LHC, triggering on jets with low 
transverse energy is not feasible. A superior trigger efficiency for the signal is obtained by using a trigger on missing transverse energy utilising only calorimeter information\cite{Casadei:1327653} (a full description of the ATLAS trigger system is given in \cite{:2008zzm}). Using an $\MET$-based trigger is possible since $R$-hadrons would typically deposit only a small fraction of their energy as they propagate through the ATLAS calorimeters. The trigger threshold applied is $\MET$  = 40\,\si{\giga\electronvolt} which gives an efficiency ranging from approximately 15\% for a gluino-mass of 100\,\si{\giga\electronvolt}{} to 32\% for a 600\,\si{\giga\electronvolt}{} mass. The missing transverse energy trigger is based on a level-1 trigger decision derived from coarsely segmented energy measurements, followed by a decision at the higher-level trigger based on the full granularity of the ATLAS calorimeter.

\subsection{Selection of R-hadron candidates}\label{sec:rsel}
Table~\ref{tab:cutflow} shows the cut flow of the analysis. After the trigger selection, each event is required to contain a track with a transverse momentum greater than 10\,\GeV. This track must be matched either to a muon reconstructed in the muon spectrometer or to a cluster in the tile calorimeter. The track is required to have MIP-compatible energy depositions in the calorimeter. Such an event is referred to in the table as a \emph{Candidate}. Each event is required to contain at least one good primary vertex, to which at least three tracks are associated.
Only tracks in the central region ($|\eta|<1.7$) are considered.
This matches the acceptance of the tile calorimeter. To ensure well measured kinematics, track quality requirements are made: the track must have at least two hits in the pixel detector, at least six hits in the silicon-strip Semiconductor Tracker, and at least six associated hits in the Transition Radiation Tracker (TRT). Jet candidates are reconstructed using the anti-$k_{\mathrm{t}}$ jet clustering algorithm \cite{Cacciari:2008gp,Cacciari:2005hq} with a distance parameter of 0.4. In order to suppress backgrounds from jet production, the distance in $\eta$-$\phi$ space between the candidate and any jet with $\ET \ge 40$\,\si{\giga\electronvolt}{} must be greater than $\Delta R=\sqrt{(\Delta\eta)^2+ (\Delta\phi)^2}=0.5$. Finally, the measured transverse momentum of the candidate must be greater than 50\,\si{\giga\electronvolt}.



\begin{table}[!htbp]
\caption{Observed and expected event yields at different steps of the data selection procedure. The individual rows of the table correspond to the stages in the cut flow as defined in the text. The rows denoted {\it Mass preselection} and {\it Final selection} indicate the number of events having at least one candidate with a mass estimate from both subsystems and passing the final mass cuts, respectively. These selections are defined in Section~\ref{sec:mass}. In addition to data and background, predictions from the signal simulations are shown. Predicted yields are scaled to the integrated luminosity of the data sample.
\label{tab:cutflow}}
~\\

\begin{scriptsize}

\begin{tabular}{lrrrrrrr}
\hline
{\bf Cut level} 		& {\bf Data  } 	& {\bf Background } 	& {\bf 300 GeV $\tilde{g}$ } 	& {\bf 500 GeV $\tilde{g}$ } 	& {\bf 600 GeV $\tilde{g}$ } 	& {\bf 200 GeV $\tilde{t}$ } 	& {\bf 200 GeV $\tilde{b}$ } \\
\hline
No cuts			& - 			& - 				&  $2.13\times10^3$ 		&    80.4 					&    21.8 					&   405 					&   405 \\
Trigger			& - 			& - 				&   616 					&    25.6 					&     6.96 					&   109 					&   108 \\
Candidate		& 75466 	& $68.0\times10^{3}$ 	&   416 					&    17.6 					&     4.80 					&    87.4 					&    67.9 \\
Vertex			& 75461 	& $68.0\times10^{3}$ 	&   416 					&    17.6 					&     4.80 					&    87.4 					&    67.9 \\ 
$|\eta| < $1.7		& 64618 	& $60.5\times10^{3}$ 	&   364 					&    15.7 					&     4.32 					&    75.2 					&    56.8 \\
Track quality		& 59872 	& $58.1\times10^{3}$ 	&   355 					&    15.3 					&     4.20 					&    73.3 					&    54.9 \\
$\Delta R > 0.5$	& 49205 	& $49.4\times10^3$ 		&   349 					&    15.1 					&     4.13 					&    72.7 					&    54.5 \\
$p_{\mathrm{T}} > 50$ GeV &  5116 &  $6.56\times10^3$ &   330 					&    14.5 					&     3.95 					&    68.9 					&    50.0 \\ \hline
Mass preselection	&    36 		&    56.0 			&   184 					&     9.70 					&     2.75 					&    32.6 					&    18.9 \\
Final selection		&    - 			&    - 				&   173 					&     9.17 					&     2.62 					&    30.6 					&    17.5 \\ 
\hline
\end{tabular}

\end{scriptsize}

\end{table}



\begin{figure}[htbp]
\ \!\!\!\!\!\!\!\!
\includegraphics[width=0.53\textwidth]{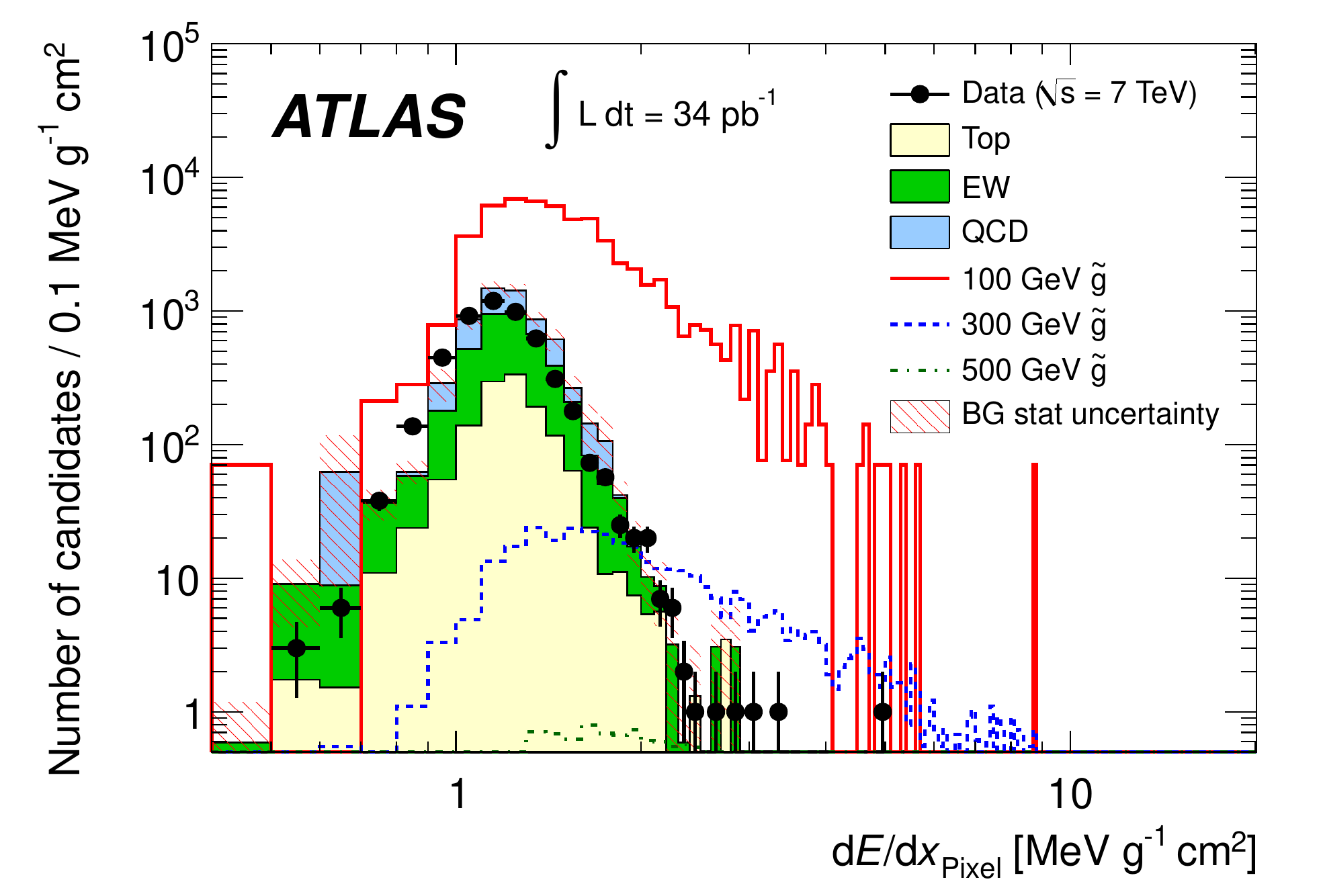}\!\!\!\!\!
\includegraphics[width=0.53\textwidth]{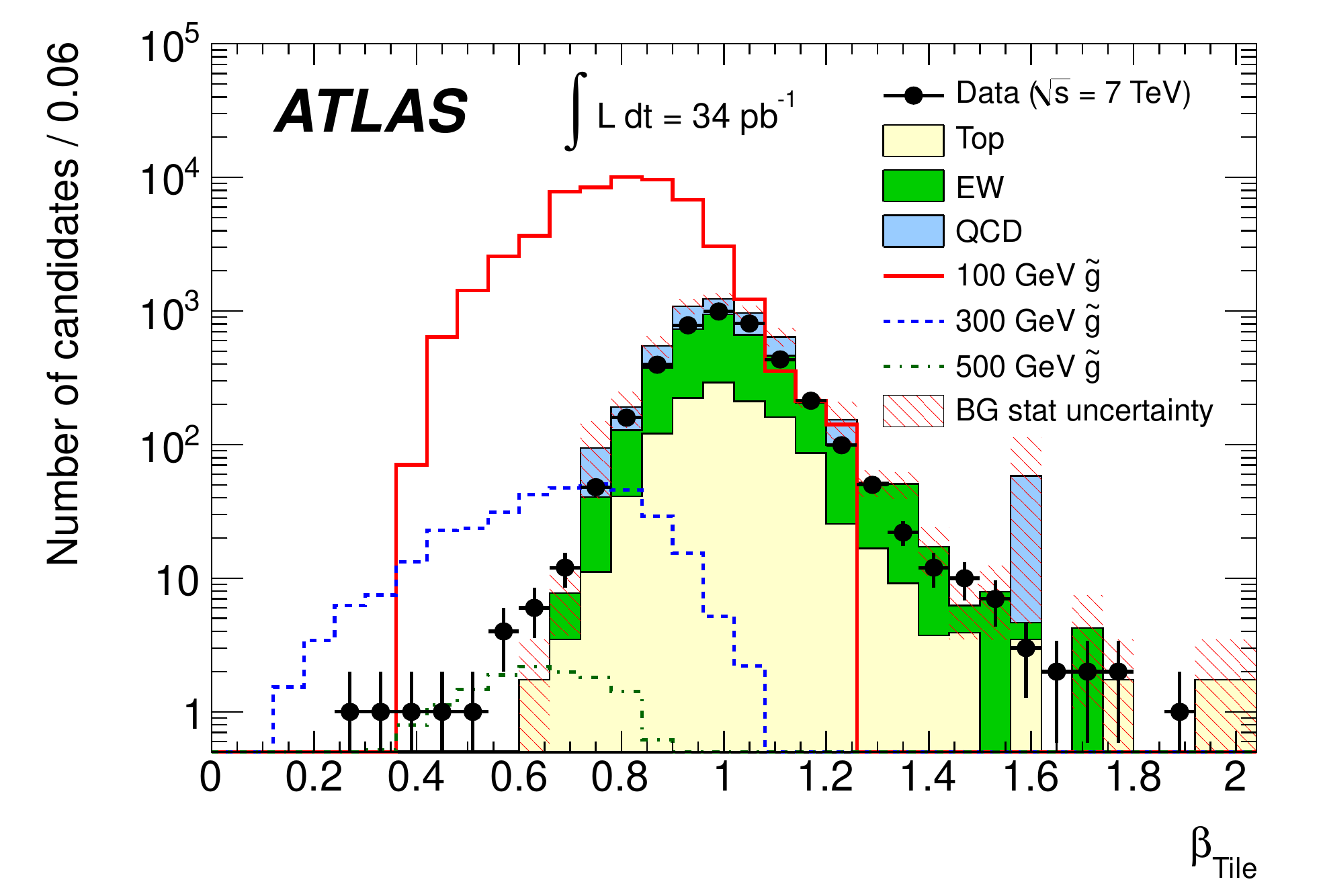}
\caption{
Distributions of \dedxp{} (left) and ${\beta}_{\mathrm{Tile}}$ (right) in data after the transverse momentum selection $\pT>50$\,\si{\giga\electronvolt}. Spectra for simulated background processes are plotted for comparison. The uncertainty shown on the background is the Monte Carlo statistical uncertainty.}
\label{fig:tilebeta}
\end{figure}

After the selection, 5208 candidates in 5116 events are observed. Figure~\ref{fig:tilebeta} shows the \dedxp{} and \betat{} distributions for these candidates together with background simulations. As can  be seen, the \betat{} measurements are centred around one.  The width of the distribution, as determined by a Gaussian fit around the bulk of the data, is $\sim$0.1. Reasonable agreement between data and the background simulations is observed, although the latter calculations are not used in any quantitative way in the analysis. The expected distributions for signal particles are overlaid and scaled to the luminosity of the data by their production cross-section, illustrating the sensitivity of these observables to $R$-hadrons.

\section{Mass reconstruction}\label{sec:mass}
For each candidate, the mass is estimated by dividing its momentum by $\beta\gamma$, determined either from pixel detector ionisation or from the tile calorimeter time-of-flight.
In the pixel detector, the following simplified Bethe-Bloch equation gives a good description of the relation between the most probable value (${\cal{M}}_{\mathrm{d}E\over \mathrm{d}x}$) of {\dedxp} and $\beta\gamma$ in the range relevant to this analysis (\mbox{$0.2< \beta\gamma < 1.5$}):

\begin{equation}
   {\cal{M}}_{\mathrm{d}E\over \mathrm{d}x}(\beta) = {p_1\over \beta^{p_3}}\ln(1+(p_2\beta\gamma)^{p_5})-p_4
  \label{eq:bbfun}
\end{equation}

To find $\beta$, and hence a mass estimate, this equation must be solved for $\beta$, identifying the measured \dedxp{} with ${\cal{M}}_{\mathrm{d}E\over \mathrm{d}x}$. This requires the {\dedxp} value to be above that of a MIP. The parameters $p_1$--$p_5$ in Equation~\ref{eq:bbfun} are determined from fits to SM particles with well-known masses and ionisation properties, $p$, $K$ and $\pi$\cite{dedxnote}, and provide a relative {\dedxp} resolution of about 10\% in the asymptotic region ($\beta\gamma > 1.5$). To reduce the backgrounds further, the final selection requires that $\dedxp > 1.8\,\si{\mega\electronvolt}\rm{g}^{-1}\rm{cm}^{2}$ compared to $\dedxp\sim 1.1\,\si{\mega\electronvolt}\rm{g}^{-1}\rm{cm}^{2}$ deposited by a MIP. In the tile calorimeter, the $\beta$-values are required to be less than 1.

\begin{figure}[htbp]
\ \!\!\!\!\!\!\!\!
\includegraphics[width=0.52\textwidth]{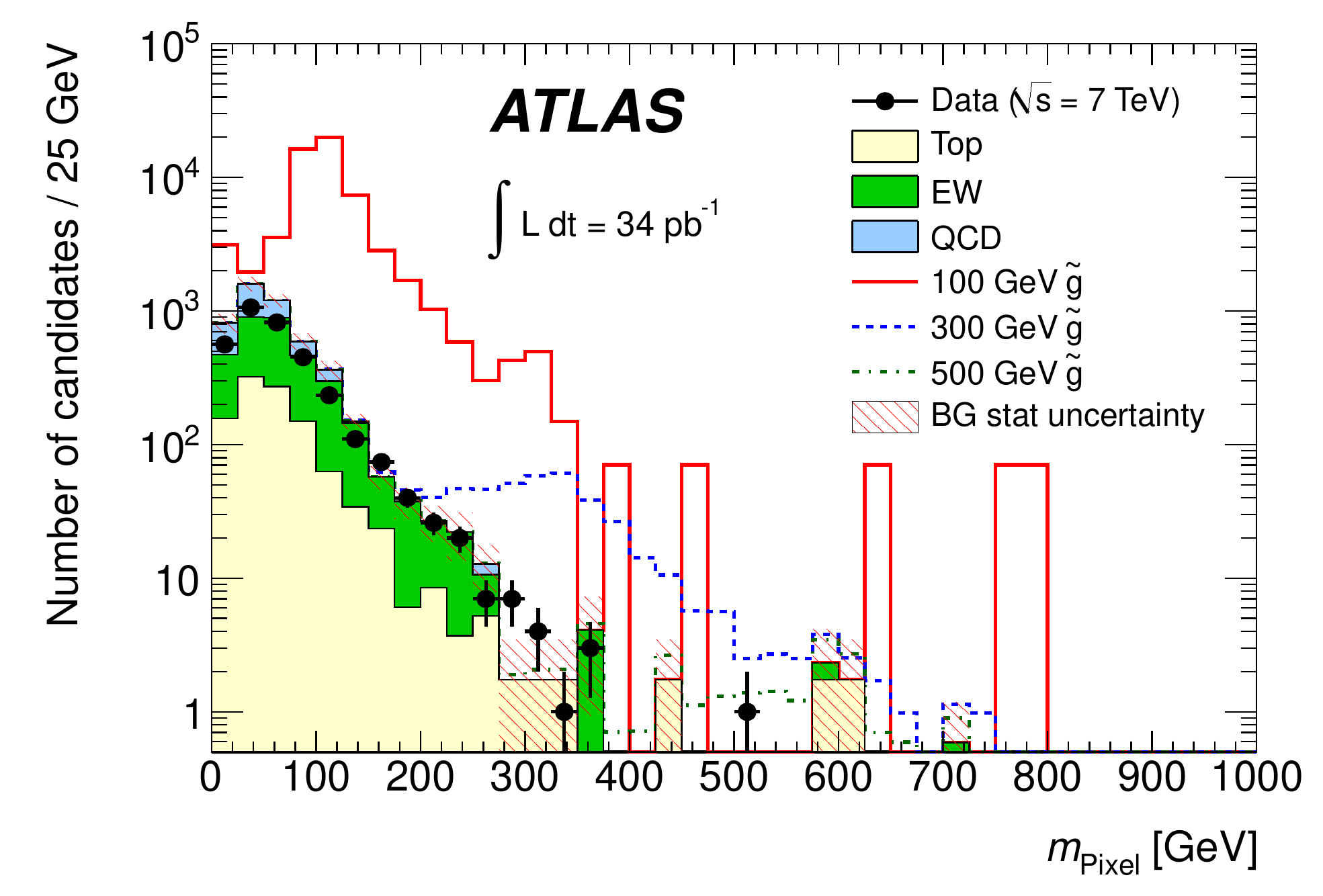}\!\!\!
\includegraphics[width=0.52\textwidth]{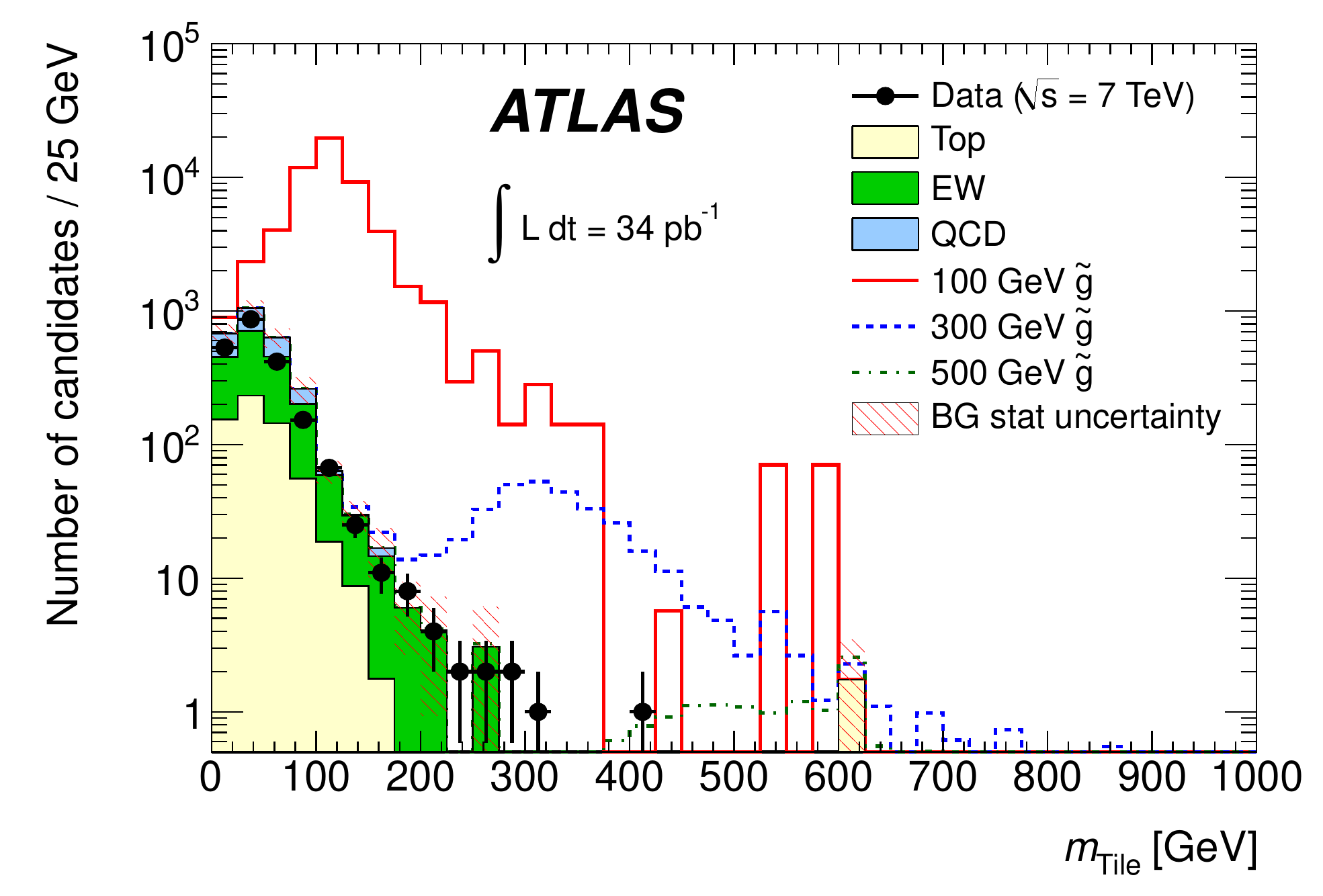}
\caption{Mass estimated by the pixel detector (left) and the tile calorimeter (right). To obtain a mass estimate, a cut of $\dedxp>1.1\,\si{\mega\electronvolt}\rm{g}^{-1}\rm{cm}^{2}$ is imposed for the pixel detector distribution. This is a looser cut than used in the analysis itself. For the tile calorimeter, the requirement is that $\betat<1$.}
\label{fig:allmass}
\end{figure}

The pixel detector and the tile calorimeter provide independent measurements from which the mass of the SMP candidate can be estimated. Making requirements on both mass estimates is a powerful means to suppress the tails in the individual distributions arising from instrumental effects. In Figure~\ref{fig:allmass} the estimated mass distributions based on \dedxp{} and ${\beta}_{\mathrm{Tile}}$ are shown
after the 50\,\si{\giga\electronvolt}{} transverse momentum cut of the event selection. In contrast to the other figures in this paper, the signal distributions are stacked on top of the background to illustrate the total expected spectra for the signal+background scenarios.

To establish signal regions for each mass hypothesis, the mean, $\mu$, and Gaussian width, $\sigma$, of the mass peak is determined for both the pixel detector and the tile calorimeter measurement. The signal region is then defined to be the region above the fitted mean minus twice the width  (i.e.\ $m_{\mathrm{Pixel}} > \mu_{\mathrm{Pixel}} - 2\sigma_{\mathrm{Pixel}}$ for the mass as estimated by the pixel detector and $m_{\mathrm{Tile}} > \mu_{\mathrm{Tile}} - 2\sigma_{\mathrm{Tile}}$ for the mass as estimated by the tile calorimeter).
The final signal region is defined by applying both of the individual mass requirements.

\section{Background estimation}\label{sec:bgest}
Rather than relying on simulations to predict the tails of the \dedxp{} and \betat{} distributions, a data-driven 
method is used to estimate the background.
No significant correlations between the measurements of momentum, \dedxp, and \betat{} are observed. This is exploited to estimate the amount of background arising from instrumental effects.
Estimates for the background distributions of the mass estimates are obtained by combining random momentum values (after the kinematic cuts defined above) with random measurements of \dedxp{} and \betat.
The sampling is performed from candidates passing the kinematic cuts defined in Section \ref{sec:rsel} for the case of \betat{}, while \dedxp{} is extracted from a sample fulfilling $10<p_T<20\,\si{\giga\electronvolt}$.

\begin{figure}[htbp]
\ \!\!\!\!\!\!\!\!
\includegraphics[width=0.525\textwidth]{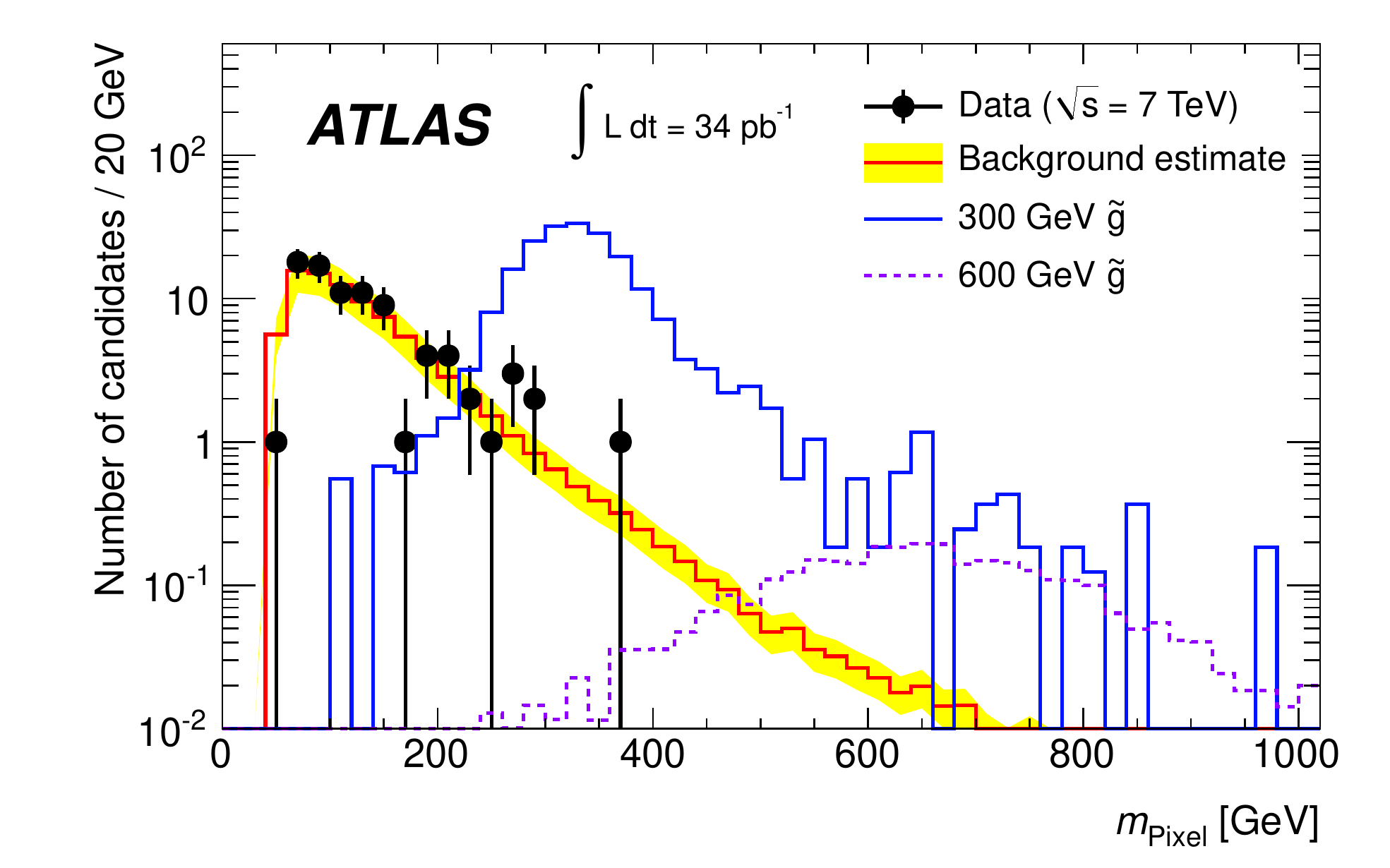}\!\!\!\!\!\!
\includegraphics[width=0.525\textwidth]{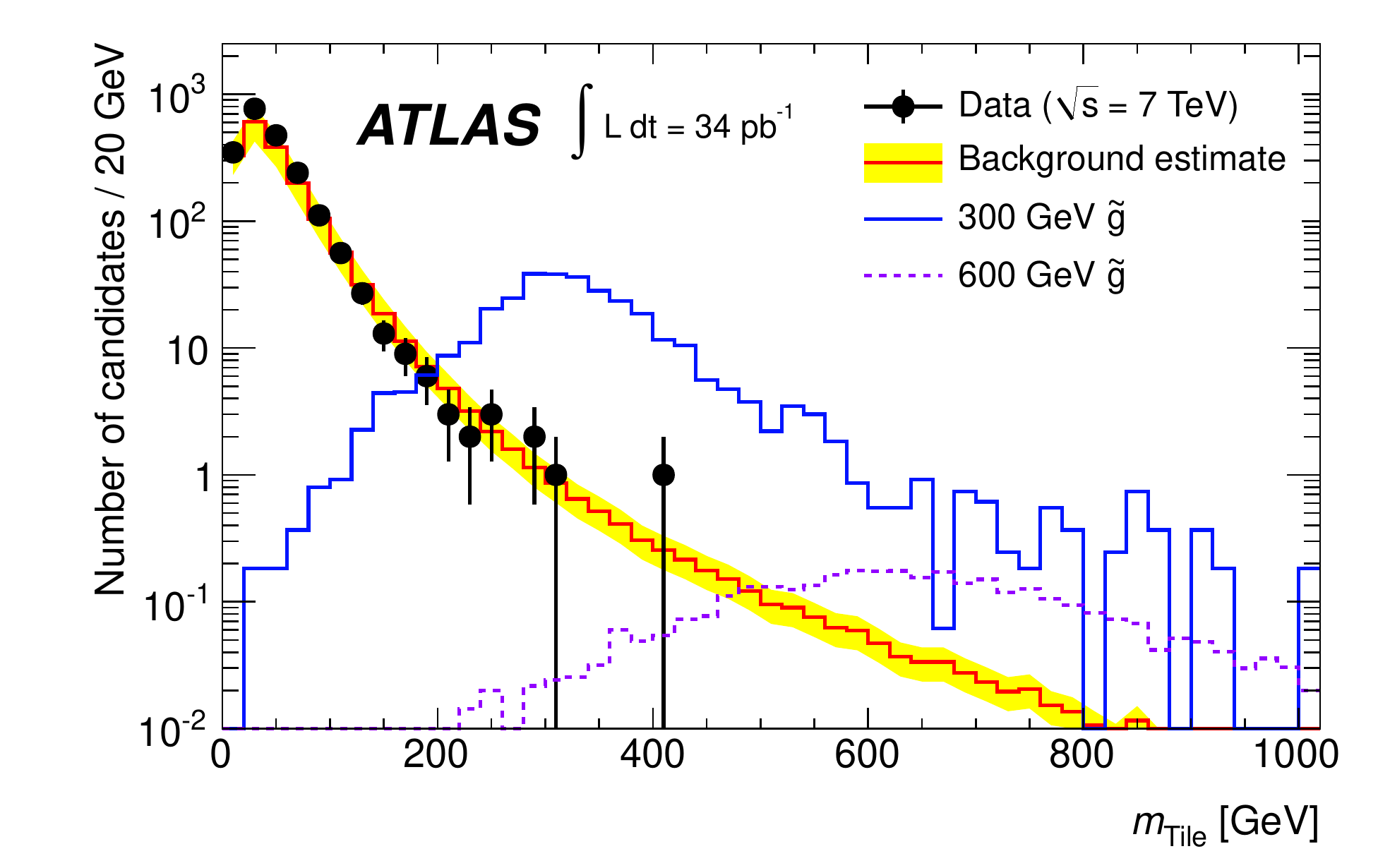}
\caption{Background estimates for the pixel detector (left) and the tile calorimeter (right).
Signal samples are superimposed on the background estimate. The total systematic uncertainty of the background estimate is indicated by the error band.}
\label{fig:bkg-est}
\end{figure}

The process is repeated many times to reduce fluctuations and the resulting estimates are normalised to match the number of events in data. The resulting background estimates can be seen in Figure~\ref{fig:bkg-est} for the pixel detector (requiring $\dedxp > 1.8\,\si{\mega\electronvolt}\rm{g}^{-1}\rm{cm}^{2}$) and the tile calorimeter (requiring $\betat<1$) separately. As can be seen from the figures, there is a good overall agreement between the distribution of candidates in data and the background estimate. The expected background at high mass is generally small.

Combining the pixel detector and the tile calorimeter mass estimates as described in Section \ref{sec:mass} further reduces
the background while retaining most of the expected signal. In contrast to the individual background estimates shown in Figure \ref{fig:bkg-est}, the combined background is obtained by combining one random momentum value with random measurements of both \dedxp{} and $\betat$.
The agreement between the distribution of candidates in data and the background estimate is good. This is seen in 
Table~\ref{tab:thresh}, which contains the event yields in the signal regions defined in Section \ref{sec:mass} for the gluino signal, for the estimated background and for real data. 
The table also contains the means and the widths of the estimated mass distributions, which are used to determine the signal regions, as described in Section~\ref{sec:mass}. Using combined data, there are no events containing a candidate with mass greater than 100\,\si{\giga\electronvolt}. There are five candidates observed for the 100\,\si{\giga\electronvolt} mass hypothesis, for which the mass window extends to values less than 100\,\si{\giga\electronvolt}. 

{
\begin{table}[!htbp]
\centering
\small
\caption{Expected number of signal and background events for the pixel detector and the tile calorimeter separately and combined for gluino mass hypotheses between 100 and 700\,\si{\giga\electronvolt}.
The fitted means and widths of the estimated mass distributions are shown on the left. 
To the right of the vertical line, the number of signal and estimated background events are shown in the relevant signal regions, along with the number of events observed in data. Systematic uncertainties are discussed in Section~\ref{sec:sys}.
\label{tab:thresh}}
\hspace{1.5cm}
\begin{tabular}{crrrr|rrrrrrrr}
\hline
Nominal&$\mu_{\mathrm{Pixel}}$&$\sigma_{\mathrm{Pixel}}$&$\mu_{\mathrm{Tile}}$&$\sigma_{\mathrm{Tile}}$&\multicolumn{3}{c}{No.\ of signal cand.\ ($\tilde{g}$)}&\multicolumn{3}{c}{Est.\ no. of bkg.\ cand.} &$N_{Data}$\\
mass~(\si{\giga\electronvolt})&(\si{\giga\electronvolt})&(\si{\giga\electronvolt})&(\si{\giga\electronvolt})&(\si{\giga\electronvolt})&Pixel &Tile&Comb.&Pixel &Tile&Comb.&Comb.\\ \hline
100 &107 & 10   &109 & 19   &15898 &49300  &13912  & 61 &330 &  5.4  & 5 \\
200 &214 & 24   &211 & 36   &1417  & 2471  &1235   & 19 &61 &  0.87   & 0 \\
300 &324 & 40   &315 & 56   &202   & 304   &173    &  6.5 &17 &  0.22   & 0 \\
400 &425 & 67   &415 & 75   & 43   & 57    & 37    &  3.4 &7.2 &  0.082  & 0 \\
500 &533 & 94   &513 &106   & 11   & 13    & 9.2   &  1.82&4.4 &  0.044  & 0 \\
600 &641 &125   &624 &145   &  3.1 &   3.5 & 2.6   &  1.08&3.2 &  0.028  & 0 \\
700 &727 &149   &714 &168   &  0.99 &  1.07& 0.84  &  0.74&2.1 & 0.018  & 0 \\
\hline
\end{tabular}
\end{table}
}

\section{Systematic uncertainties and checks}\label{sec:sys}

A number of sources of systematic uncertainties are investigated. This section describes uncertainties arising due to the limited accuracy of  theory calculations used in this work together with experimental uncertainties affecting the signal efficiency and background estimate. 

Uncertainties due to the limited accuracy of perturbative QCD calculations are studied in the following way. The production cross-section from {\sc Prospino} is calculated using the sparticle mass as the renormalisation scale with uncertainties estimated by varying the renormalisation and factorisation scales upward and downward by a factor of two in accordance with Ref.~\cite{Beenakker:1996ch}. This leads to a broadly mass-independent uncertainty of $\sim $15\% in the event yield. A variation of less than 5\% is observed substituting the MSTW 2008 NLO PDF set~\cite{Martin:2009iq} for CTEQ 6.6. Variations of scale parameters used in {\sc Pythia} to model higher-order radiation are also performed within the range allowed by data\cite{Aad:2009wy}. This leads to an uncertainty of $\sim$10\% in the signal efficiency.

A systematic shift in the scale of the missing transverse energy in the simulation of the signal would lead to a change in trigger efficiency and hence signal acceptance. This uncertainty is estimated by varying the missing transverse energy by the corresponding scale uncertainty\cite{:2010wv}. The result is an effect of 7-13\%{} on the relative signal efficiency. Based on the difference between the trigger efficiency for data and the simulation for events containing a $W$ boson decaying muonically, a further 3-5\%{} systematic uncertainty is applied. Both of these effects depend on the mass of the signal sample, and the larger uncertainties apply to the low-mass scenarios.

Uncertainties arising from track reconstruction are also studied. To account for differences in detector alignment between the simulation and data, a smearing is applied to the track $\pT$ which describes the performance observed for high-$\pT$ muons as a function of $\eta$ and $\pT$. Doubling the smearing has a negligible effect on the predicted yields. Furthermore, to account for data/simulation differences in track reconstruction efficiency, a 2\% uncertainty on the signal yield is assumed \cite{Aad:2010rd}.

Only calorimeter cells measuring an energy above a threshold of 500\,\si{\mega\electronvolt} are used in the calculation of \betat. To study the impact of this threshold on the efficiency of the measurement, the tile calorimeter cell energy scale is varied by~$\pm 5\%$~\cite{ATL-CAL-PUB-2010-001} leading to a small ($\le 1\%$) effect on  the predicted yields of $R$-hadrons which fall into the individual signal regions. The predicted cell time distributions are smeared to match the data. To evaluate the sensitivity of the signal yield to this smearing, the smearing is applied twice, and the impact is seen to be less than 1\%.

To estimate the effects of an imperfect description of the \dedxp{} resolution by the simulation, individual values of \dedxp{} are smeared according to a Gaussian function with width 5\%\cite{dedxnote}. Furthermore, to study possible effects due to a global \dedxp{} scale uncertainty, the scale is shifted by $\pm 3\%$. These variations are motivated by observed differences between data and Monte Carlo simulations and they change the predicted number of events passing the signal selections by less than 1\%.

Adding the above errors in quadrature together with an 11\% uncertainty from the luminosity measurement~\cite{Aad:2011dr}, a total 
systematic uncertainty of 17-20\% on the signal event yield is estimated, where the larger uncertainty applies to the low-mass 
scenarios. The systematic uncertainty on the background estimate is found to be 30\%. This arises from contributing uncertainties in the \dedxp and \betat distributions (25\%) and the use of different methods to determine the absolute normalisation of the background prediction (15\%).

As a final cross-check of the consistency of the analysis, the TRT was used. The TRT is a straw-based gas detector, and the time 
in which any signal exceeds the threshold is read out. This time provides an estimate of continuous energy loss and is usable for particle identification~\cite{Abat:1900zz}. The measurement is similar to (but independent of) the pixel detector time-over-threshold
measurement, from which \dedxp{} is based. No deviations from backgrounds expectations are observed, and the TRT thus provides an additional confirmation that no signal was missed.

\section{Exclusion limits}
\begin{figure}[htbp]
\vspace{-0.3cm}
  \begin{center}
    \includegraphics[width=0.7\textwidth]{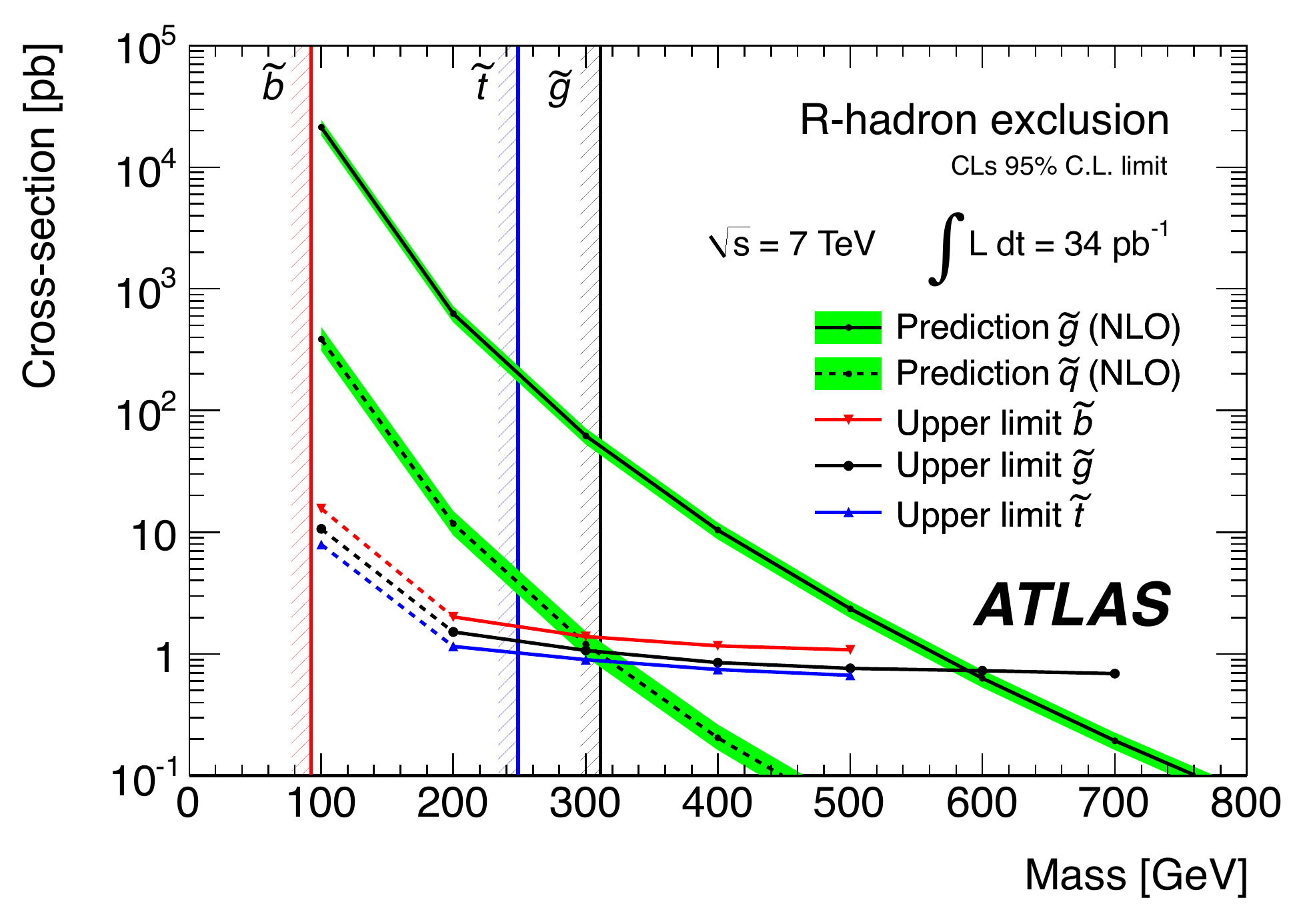}
  \end{center}
  \vspace{-1.0cm}
  \caption{Cross-section limits at 95\%{} CL as a function of sparticle mass. Since five candidate events are observed for the 
mass windows used for the 100\,\si{\giga\electronvolt} mass hypotheses, the mass points between 100 and 200\,\si{\giga\electronvolt} are connected with a dotted line. This indicates that fluctuations in the excluded cross-section will occur. The mass limits quoted in the text are inferred by comparing the cross-section limits with the model predictions. Systematic uncertainties from the choice of PDF and the choice of renormalisation and factorisation scales are represented as a band  in the cross-section curves. Previous mass limits are indicated by shaded vertical lines for sbottom (ALEPH), stop (CDF) and gluino (CMS).}
\label{fig:limmass}
\end{figure}

Given an expected cross-section as calculated by {\sc Prospino} and our computed efficiency, the expected number of signal events as a function of mass is determined and a lower limit on the $R$-hadron mass using the $CL_s$ method \cite{Read:451614} is calculated.  The results for the signal models defined in Section \ref{sec:mc} are summarised in Figure~\ref{fig:limmass}.

The observed 95\%{} CL limits are 294\,\si{\giga\electronvolt}{} for sbottom $R$-hadrons and 309\,\si{\giga\electronvolt}{} for stop $R$-hadrons, while the lower limit for the mass of a hadronising gluino is 586\,\si{\giga\electronvolt}. These limits include the systematic uncertainties on the signal cross-section and efficiency, as well as on the data-driven background estimate, as described above. Evaluating the mass limits for gluino $R$-hadrons using the triple-Regge based model and bag-model calculation of Ref.~\cite{Farrar:2010ps}, gives 566 and 562\,\si{\giga\electronvolt}{} respectively. The lower mass limits from ATLAS are shown in Figure~\ref{fig:limmass} and compared with earlier results from ALEPH~\cite{Heister:2003hc} (sbottom) , CDF~\cite{Aaltonen:2009ke} (stop), and CMS~\cite{Collaboration:2011ts} (gluino). The ATLAS limits have a higher mass reach than those obtained from the previous searches.

\section{Summary}
A search has been performed for slow-moving squark- (stop and sbottom) and gluino-based $R$-hadrons, pair-produced in proton-proton collisions at 7 TeV centre-of-mass energy at the ATLAS detector at the LHC. Candidate $R$-hadrons were sought which left a high transverse momentum track associated with energy depositions in the calorimeter. Observables sensitive to $R$-hadron speed (ionisation energy loss and time-of-flight) were used to suppress backgrounds and allow the reconstruction of the candidate mass. The influence of the scattering of $R$-hadrons in matter on the search sensitivity was studied using a range of phenomenological scattering models. At 95\% confidence level the most conservative lower limits on the masses of stable sbottoms, stops and gluinos are 294, 309, and 562\,\si{\giga\electronvolt}, respectively. Each of these limits are the most stringent to date.

\section{Acknowledgements}

We wish to thank CERN for the efficient commissioning and operation of the LHC during this initial high-energy data-taking period as well as the support staff from our institutions without whom ATLAS could not be operated efficiently. We would also like to thank Torbj\"orn Sj\"ostrand and Tilman Plehn for their assistance in the preparation of the theory calculations used in this work.

We acknowledge the support of ANPCyT, Argentina; YerPhI, Armenia; ARC, Australia; BMWF, Austria; ANAS, Azerbaijan; SSTC, Belarus; CNPq and FAPESP, Brazil; NSERC, NRC and CFI, Canada; CERN; CONICYT, Chile; CAS, MOST and NSFC, China; COLCIENCIAS, Colombia; MSMT CR, MPO CR and VSC CR, Czech Republic; DNRF, DNSRC and Lundbeck Foundation, Denmark; ARTEMIS, European Union; IN2P3-CNRS, CEA-DSM/IRFU, France; GNAS, Georgia; BMBF, DFG, HGF, MPG and AvH Foundation, Germany; GSRT, Greece; ISF, MINERVA, GIF, DIP and Benoziyo Center, Israel; INFN, Italy; MEXT and JSPS, Japan; CNRST, Morocco; FOM and NWO, Netherlands; RCN, Norway;  MNiSW, Poland; GRICES and FCT, Portugal;  MERYS (MECTS), Romania;  MES of Russia and ROSATOM, Russian Federation; JINR; MSTD, Serbia; MSSR, Slovakia; ARRS and MVZT, Slovenia; DST/NRF, South Africa; MICINN, Spain; SRC and Wallenberg Foundation, Sweden; SER,  SNSF and Cantons of Bern and Geneva, Switzerland;  NSC, Taiwan; TAEK, Turkey; STFC, the Royal Society and Leverhulme Trust, United Kingdom; DOE and NSF, United States of America.  

The crucial computing support from all WLCG partners is acknowledged gratefully, in particular from CERN and the ATLAS Tier-1 facilities at TRIUMF (Canada), NDGF (Denmark, Norway, Sweden), CC-IN2P3 (France), KIT/GridKA (Germany), INFN-CNAF (Italy), NL-T1 (Netherlands), PIC (Spain), ASGC (Taiwan), RAL (UK) and BNL (USA) and in the Tier-2 facilities worldwide.

\newpage
\bibliographystyle{elsarticle-num}
\bibliography{atlaspaper}


\newpage
\begin{flushleft}
{\Large The ATLAS Collaboration}

\bigskip

G.~Aad$^{\rm 48}$,
B.~Abbott$^{\rm 111}$,
J.~Abdallah$^{\rm 11}$,
A.A.~Abdelalim$^{\rm 49}$,
A.~Abdesselam$^{\rm 118}$,
O.~Abdinov$^{\rm 10}$,
B.~Abi$^{\rm 112}$,
M.~Abolins$^{\rm 88}$,
H.~Abramowicz$^{\rm 153}$,
H.~Abreu$^{\rm 115}$
% ...
\bigskip

% ...
$^{10}$ Institute of Physics, Azerbaijan Academy of Sciences, Baku, Azerbaijan\\
$^{11}$ Institut de F\'isica d'Altes Energies and Universitat Aut\`onoma  de Barcelona and ICREA, Barcelona, Spain\\
% ...
$^{48}$ Fakult\"{a}t f\"{u}r Mathematik und Physik, Albert-Ludwigs-Universit\"{a}t, Freiburg i.Br., Germany\\
$^{49}$ Section de Physique, Universit\'e de Gen\`eve, Geneva, Switzerland\\
% ...
$^{88}$ Department of Physics and Astronomy, Michigan State University, East Lansing MI, United States of America\\
% ...
$^{111}$ Homer L. Dodge Department of Physics and Astronomy, University of Oklahoma, Norman OK, United States of America\\
$^{112}$ Department of Physics, Oklahoma State University, Stillwater OK, United States of America\\
% ...
$^{115}$ LAL, Univ. Paris-Sud and CNRS/IN2P3, Orsay, France\\
% ...
$^{118}$ Department of Physics, Oxford University, Oxford, United Kingdom\\
% ...
$^{153}$ Raymond and Beverly Sackler School of Physics and Astronomy, Tel Aviv University, Tel Aviv, Israel\\
% ...
\end{flushleft}

\end{document}